\documentclass [12pt,showkeys, showpacs]  {revtex4}
\begin{document}
\title[Short Title]{Secure Quantum Secret Sharing Based on Reusable GHZ States as Secure Carriers }
\author{Jie SONG and Shou ZHANG\footnote{E-mail: szhang@ybu.edu.cn} }
\affiliation{Department of Physics, College of Science, Yanbian
University, Yanji, Jilin 133002, PR China}
\begin{abstract}
We show a potential eavesdropper can eavesdrop
 whole secret information when the legitimate users use secure carrier
 to encode and decode classical information repeatedly in the protocol
 [proposed in Bagherinezhad S and  Karimipour V 2003 Phys. Rev. A
\textbf{67} 044302]. Then we present a revised quantum secret
sharing  protocol by using Greenberger-Horne-Zeilinger state as
secure carrier. Our protocol can resist Eve's attack.
 \end{abstract}
\pacs {03.67. Hk, 03.65. Ud} \keywords{GHZ state, controlled-Not
operation, Hadamard operation}
\maketitle

Quantum secret sharing (QSS) is a generalization of classical secret
sharing to a quantum scenario. The basic idea of QSS is: After Alice
distributes her secret message to the receivers (Bob, Charlie,
\ldots), all the receivers can obtain the secret message
simultaneously by collaboration. Since Hillery {\it et al.} \cite{4}
proposed the pioneering QSS work, QSS has attracted a widespread
attention and progressed quickly \cite{5,6,7,8,10,11,12,13} over the
past several years. In 2001, Guo {\it et al.} \cite{9} proposed a
quantum key distribution protocol by using quantum key to encode and
decode classical information. The quantum key can be used
repeatedly. Based on Guo's protocol, Karimipour {\it et al.}
\cite{1} proposed a QSS protocol by using
Greenberger-Horne-Zeilinger (GHZ) states as secure carriers.

Recently, Gao {\it et al.} \cite{2} present an eavesdropping
protocol by which Eve can obtain odd numbered data bits of
Karimipour's protocol \cite{1}. Karimipour \cite{3} shows that the
quantum bits in odd rounds can be discarded by legitimate users.
Only if Eve can get the whole information, this protocol can be
proved insecure. In this paper, we propose another eavesdropping
method to the protocol in Ref. \cite{1}. Eve can get the whole
information without being detected by the legitimate users. Then
we present a revised QSS protocol by using GHZ state to encode and
decode classical information. Our QSS protocol can be robust
against Eve's attack.

First of all, let us give a brief review of Karimipour's protocol
\cite{1}. In the beginning, Alice, Bob and Charlie have shared a
GHZ state
$|\psi\rangle=\frac{1}{\sqrt{2}}(|0,0,0\rangle+|1,1,1\rangle)_{a,b,c}$
as carrier (used subscripts a, b and c on the states and operators
for Alice, Bob and Charlie, respectively), and the data bits Alice
wants to distributed to Bob and Charlie can be represented by
$q_1, q_3, q_5,\cdots$ (where $q=0 ~\text{or}~ 1$) in odd round
and $q_2, q_4, q_6, \cdots$ (where
$q\rightarrow\bar{q}=|q\rangle_{12}=\frac{1}{\sqrt{2}}\{|0,
q\rangle+|1, q\oplus1\rangle\}$ and ``$\oplus$'' is performed
modulo 2) in even round. In odd round, Alice entangles the state
$|q,q\rangle_{1,2}$ to the carrier by performing  controlled-NOT
(CNOT) operations $C_{a1}$, $C_{a2}$ and send them to Bob and
Charlie ($C_{a1}$ is the controlled-NOT gate and is specified by
two subscripts, the first one is the control bit, the second one
is the target bit). Bob and Charlie act on this state by the
operations $C_{b1}$, $C_{c2}$ to read independently his own bit.
In even round, Alice performs $C_{a1}$ to entangle the state
$|\overline{q}\rangle_{12}$ to the carrier. Bob and Charlie
disentangle the state $|\overline{q}\rangle_{12}$ from the carrier
by the operations $C_{b1}$, $C_{c2}$. Bob and Charlie can gain
Alice's information by collaboration. Now let us describe our
eavesdropping strategy to eavesdrop the whole information of \cite
{1}.  Eve prepares two qubits ($e_1$ and $e_2$) in state
$|0\rangle$ as her two ancilla.

(1) In the first round, according to the method in Ref. \cite{2},
Eve intercepts the first sending qubit which is denoted by
subscript 1, and then performs a controlled-NOT (CNOT) operation
$C_{1e_1}$ on this qubit and her ancilla after Alice sent the
particles. This state of these qubits can be specified by
\begin{equation}\label{e1}
|\psi^{1}\rangle=\left\{\begin{array}{ll}\frac{1}{\sqrt{2}}(|0,0,0,0,0,0\rangle+|1,1,1,1,1,1\rangle)_{a,b,c,e_1,1,2}
\ \ &(q_1 = 0), \\ \\
\frac{1}{\sqrt{2}}(|0,0,0,1,1,1\rangle+|1,1,1,0,0,0\rangle)_{a,b,c,e_1,1,2}\
\ \ &(q_1 = 1),\end{array}\right .
\end{equation}
and then Eve resends it to Bob. Here superscript 1 denotes the
first round. The value of $q_n$ denotes Alice's sending
information. Bob and Charlie can get the right information when
they disentangle the sending qubit from the carrier.

(2) In the second round, as Alice, Bob and Charlie perform a
Hadamard gate on their particles, respectively, Eve also performs
the same operation on his ancilla $e_1$. After Alice  entangles
the entangled state $\overline{q_2}$ to the carrier, she sends the
two particles (1 and 2) to Bob and Charlie, then Eve intercepts
the two sent qubits and performs two CNOT operations $C_{1e_2}$,
$C_{2e_2}$. The qubit $e_2$ is also entangled to the carrier. Eve
performs a CNOT operation $C_{e_1 1}$. She resends them to Bob and
Charlie. After Bob and Charlie performed two CNOT operations
$C_{b1}$, $C_{c2}$ to disentangle their receiving particles from
the carrier and measured the qubits 1 and 2, they cooperate to get
the value of $q_2$ and can't detect the presence of Eve, while
leaving the state
\begin{eqnarray}\label{e4}{\footnotesize
|\psi^{2}\rangle\rightarrow\left\{\begin{array}{llll}\frac{1}{2\sqrt{2}}(|0,0,0,0,0\rangle+|0,0,1,1,0\rangle+|0,1,0,1,0\rangle+|0,1,1,0,0\rangle
+|1,0,0,1,1\rangle\cr\cr~~~~+|1,0,1,0,1\rangle+|1,1,0,0,1\rangle+|1,1,1,1,1\rangle)_{a,b,c,e_1,e_2}
~~~~(q_2 = 0,q_1 = 0),\\ \\
\frac{1}{2\sqrt{2}}(|0,0,0,0,1\rangle+|0,0,1,1,1\rangle+|0,1,0,1,1\rangle+|0,1,1,0,1\rangle
+|1,0,0,1,0\rangle\cr\cr~~~~+|1,0,1,0,0\rangle+|1,1,0,0,0\rangle+|1,1,1,1,0\rangle)_{a,b,c,e_1,e_2}
~~~~(q_2 = 1,q_1 = 0),\\ \\
\frac{1}{2\sqrt{2}}(|0,0,0,0,0\rangle-|0,0,1,1,0\rangle-|0,1,0,1,0\rangle+|0,1,1,0,0\rangle
-|1,0,0,1,1\rangle\cr\cr~~~~+|1,0,1,0,1\rangle+|1,1,0,0,1\rangle-|1,1,1,1,1\rangle)_{a,b,c,e_1,e_2}
~~~~(q_2 = 0,q_1 = 1),\\ \\
\frac{1}{2\sqrt{2}}(|0,0,0,0,1\rangle-|0,0,1,1,1\rangle-|0,1,0,1,1\rangle+|0,1,1,0,1\rangle
-|1,0,0,1,0\rangle\cr\cr~~~~+|1,0,1,0,0\rangle+|1,1,0,0,0\rangle-|1,1,1,1,0\rangle)_{a,b,c,e_1,e_2}
~~~~(q_2 = 1,q_1 = 1).\end{array}\right .}
\end{eqnarray}

(3) In the third round, Eve can eavesdrop the odd round's key bit.
Firstly, Eve also performs a Hadamard gate on each of her two
ancilla, when Alice, Bob and Charlie perform a Hadamard gate on
their respective particles. The state of Alice, Bob, Charlie and
Eve can be expressed as
\begin{equation}\label{e5}{\footnotesize
|\psi^{3}\rangle\rightarrow\left\{\begin{array}{llll}\frac{1}{2}(|0,0,0,0,0\rangle+|1,1,1,1,0\rangle+|1,0,0,0,1\rangle+|0,1,1,1,1\rangle)_{a,b,c,e_1,e_2}
~~~(q_1 = 0,q_2 = 0),\\ \\
\frac{1}{2}(|0,0,0,0,0\rangle+|1,1,1,1,0\rangle-|1,0,0,0,1\rangle-|0,1,1,1,1\rangle)_{a,b,c,e_1,e_2}
~~~(q_1 = 0,q_2 = 1),\\ \\
\frac{1}{2}(|1,1,1,0,0\rangle+|0,0,0,1,0\rangle+|0,1,1,0,1\rangle+|1,0,0,1,1\rangle)_{a,b,c,e_1,e_2}
~~~(q_1 = 1,q_2 = 0),\\ \\
\frac{1}{2}(|1,1,1,0,0\rangle+|0,0,0,1,0\rangle-|0,1,1,0,1\rangle-|1,0,0,1,1\rangle)_{a,b,c,e_1,e_2}
~~~(q_1 = 1,q_2 = 1),\end{array}\right .}
\end{equation}
After Alice sent out the two qubits, Eve intercepts the two qubits
and performs four CNOT operations $C_{e_1 1}$, $C_{e_2 1}$ and
$C_{e_1 2}$, $C_{e_2 2}$. Eve makes a measurement on one of two
qubits (1 and 2) and gets the value of $q_3\oplus q_1$. After Eve
performs $C_{e_1 1}$ and $C_{e_1 2}$, she resends the two qubits
(1 and 2) to Bob and Charlie. Bob and Charlie perform $C_{b 1}$
and $C_{c 2}$ on their respective qubits. We conclude that Bob and
Charlie get correct result.

(4) In the fourth round, Alice, Bob, Charlie and Eve perform a
Hadamard gate operation on their respective qubits. As a result,
the state of whole system will be converted into Eq.~(\ref{e4}).
In this round, Eve can eavesdrop the key bit of even round without
being detected. Suppose information bit of Alice is
$q_4=|\bar{k}\rangle_{12}$\ ($k=0$\ or\ $1$), after Alice
performed a CNOT operation $C_{a 1}$ and sent the two qubits to
Bob and Charlie, Eve intercepts the two qubits and performs a CNOT
operation $C_{e_2 1}$. The two qubits have disentangled from the
carrier. Eve can perform two single particle measurements on the
two qubits (1 and 2) and get the value of $q_2\oplus q_4$.
Afterward, she prepares two qubits in the state
$|\overline{q_2\oplus q_4}\rangle_{1',2'}$ and performs CNOT
operations $C_{e_2 1'}$, $C_{e_1 1'}$ and sends them to Bob and
Charlie, respectively. Bob and Charlie act on their respective
qubits by the operations $C_{b 1'}$ and $C_{c 2'}$, so they can
get the correct information.

(5) In next odd and even rounds, Eve uses the same strategy in
step (3) and (4). So Eve can get the information ( $q_1\oplus q_3,
q_1\oplus q_5,\cdots$ and $q_2\oplus q_4, q_2\oplus q_6,\cdots$).
As long as any odd and even numbered data bits are announced by
Alice, Eve could obtain Alice's full information.

Comparing with Gao's strategy \cite{2} our strategy improves the
efficiency of eavesdropping. We can prove that the protocol
\cite{1} is insecure by our eavesdropping strategy. So we modify
this protocol \cite{1} as follows:

(1) Alice, Bob and Charlie have shared one three-particle state
$|\chi\rangle$ as carrier (the state can be expressed as
$|\chi\rangle=\frac{1}{\sqrt{2}}(|0,0,0\rangle+|1,1,1\rangle)_{a,b,c}$
). The following property will be used:
\begin{equation}\label{e6}
H\otimes H\otimes
H|\chi\rangle=|G\rangle=\frac{1}{2}(|0,0,0\rangle
+|1,1,0\rangle+|1,0,1\rangle+|0,1,1\rangle)_{a,b,c}.
\end{equation}

(2) Before Alice sends her information qubits to Bob and Charlie,
she decides if she performs a Hadamard gate on her qubit a. If
Alice performs a Hadamard gate on her qubit, she chooses two
qubits which are in two-qubit state
$q\rightarrow\bar{q}=|q\rangle_{12}=\frac{1}{\sqrt{2}}(|0,
q\rangle+|1, q\oplus1\rangle)$ for sending a classical bit. Alice
performs one single CNOT gate $C_{a 1}$ to entangle this state to
the carrier
\begin{equation}\label{e7}
|\psi\rangle=\frac{1}{2}\{(|0,\overline{q}\rangle+|1,
\overline{q\oplus1}\rangle)\otimes|0,0\rangle+(|0,\overline{q}\rangle
-|1,\overline{q\oplus 1}\rangle)\otimes|1,1\rangle\}_{a,1,2,b,c},
\end{equation}
and sends the two qubits to Bob and Charlie. After Bob and Charlie
have received the two qubits, Alice informs that she did a
Hadamard gate. Then Bob and Charlie also perform a hadamard gate
accordingly. This state of these qubits will be specified by
\begin{equation}\label{e8}
|\psi\rangle=\frac{1}{2}\{|0,\overline{q}\rangle\otimes(|0,0
\rangle+|1,1\rangle)+|1,\overline{q+1}\rangle\otimes(|0,1
\rangle+|1,0\rangle)\}_{a,1,2,b,c}.
\end{equation}
Bob and Charlie perform two CNOT operations $C_{b1}$ and $C_{c2}$
on their respective qubits to disentangle qubits 1 and 2.

(2') If Alice does't perform a Hardamard gate on his qubit a, she
will choose two single particles which are in the state
$|q_1\rangle$ and $|q_2\rangle$ ($q_1$, $q_2$ $\in$ (0,1)). The
classical bit is the value of $q_1\oplus q_2$. Alice performs one
CNOT gate $C_{a1}$ or $C_{a2}$ to entangle one of the two
particles to the carrier. Suppose Alice performs $C_{a1}$. The
state of these qubits can be expressed as
\begin{equation}\label{e9}
|\psi\rangle=\frac{1}{\sqrt{2}}\{(|0,0,0,q_1\rangle+|1,
1,1,q_1\oplus1\rangle)\otimes|q_2\rangle\}_{a,b,c,1,2}.
\end{equation}
Alice sends the two qubits to Bob and Charlie. After Bob and
Charlie have received the two qubits, Alice informs that she did
not perform a Hadamard gate and performed a CNOT gate $C_{a1}$.
Then Bob perform $C_{b1}$ to disentangle his receiving qubit 1.

(3) Bob and Charlie performs single particle measurement on their
respective particles (1 and 2). Only one of them publicly
announces the measurement result, and the other will add the
publicly announced result to his own measurement result to get
Alice's information.

(4) In next round, Alice know the three-particle state is in
$|\chi\rangle$ or $|G\rangle$ according to the number of Hadamard
gate she has performed. Alice, Bob and Charlie do similar
operations from step (2) to (4).

(5) At last, Alice publicly announces part of classical bits
through a classical channel. They compare the results and analyze
the error rate. If the error rate is reasonably low, they can
believe the security of the process. Otherwise they terminate the
QSS and start next one form step (1).

Now, let us discuss the security of the revised protocol. For
convenience, we call Gao's attack strategy in \cite{2} A1, and
call our attacking strategy A2. Suppose Eve intends to use A1 to
attack. Eve intercepts one of the two qubits 1 after Alice sent
the qubits. Eve prepares a qubit in the state $|0\rangle$ as her
ancilla e. She performs a CNOT operation $C_{1e}$. Eve doesn't
know that the qubits sent by Alice were in single particle state
or two-qubit state. If the qubits sent by Alice are in two-qubit
state (i.e., $|\overline{0}\rangle$ ), the state of whole system
will be converted into
\begin{eqnarray}\label{e10}
|\psi\rangle=\frac{1}{2}\{|0\rangle\otimes(|0,0,0\rangle
+|1,1,1\rangle)\otimes |0,0\rangle+|1\rangle\otimes(|1,0,1\rangle
+|0,1,0\rangle)\otimes |1,1\rangle\}_{a,1,2,e,b,c,}.
\end{eqnarray}
After Bob and Charlie received Alice's sending qubits, Alice
announces that she performed a Hadamard gate. Bob and Charlie also
perform a Hadamard gate on the qubits b and c. Then they perform a
CNOT operation $C_{b1}$ and $C_{c2}$ and make a single particle
measurement on their qubits (1 and 2), respectively. The state of
these qubits will be converted into one of the two states:
\begin{equation}\label{e11}
|\psi\rangle=\left\{\begin{array}{ll}\frac{1}{2}(|0,0,0,0\rangle+
|0,1,1,1\rangle+|1,0,1,0\rangle+|1,1,0,1\rangle)_{a,b,c,e},
\ \ & \\ \\
\frac{1}{2}(|0,0,0,1\rangle+
|0,1,1,0\rangle+|1,0,1,1\rangle+|1,1,0,0\rangle)_{a,b,c,e}.\ \ \
&\end{array}\right .
\end{equation}
In next round, when Alice sends single particles as information
qubits to Bob and Charlie, the eavesdropping behavior makes error
in comparing the data bits jointly received by Bob and Charlie
with those sent by Alice. If Alice sends single particles as
information qubits in the beginning, Eve performs a CNOT operation
$C_{1e}$ on the intercepted qubit and her ancilla. Eve might
entangle his ancilla to the carrier. But in next round Alice
doesn't announce if she has performed a hadamard gate until Bob
and Charlie received her sending qubits. Owing to Eve doesn't know
how to operate the intercepted qubit, she inevitably introduces
error. So Eve can't get any useful information and her
eavesdropping behavior will be detected by the legitimate users.

Suppose Eve wants to entangle his ancilla to the carrier by using
the method of even round in A2 to attack. After Alice sent his two
information bits to Bob and Charlie, Eve intercepts the two
sending qubits and  performs two CNOT operations $C_{1e}, C_{2e}$
on the two particles and her ancilla. Eve's aim is to entangle the
state of his ancilla to the state $|G\rangle$. When Alice sends
single particles as information qubits and performs a CNOT gate
$C_{a1}$, the state of whole system can be expressed as
\begin{equation}\label{e12}
|\phi\rangle=\frac{1}{\sqrt{2}}\{(|0,0,0,q_1,q_1\oplus q_2\rangle
+|1,1,1,q_1\oplus 1,q_1\oplus
q_2\oplus1\rangle)\otimes|q_2\rangle\}_{a,b,c,1,e,2}.
\end{equation}
When the two qubits sent by Alice are in two-qubit state, Eve can
also entangle her ancilla to the carrier. Then the state of these
qubits can be as follows
\begin{equation}\label{e13}
|\phi'\rangle=\frac{1}{2}\{(|0,0,0,\overline{q},q\rangle
+(|1,1,0,\overline{q\oplus1},q\oplus1\rangle+(|1,0,1,\overline{q\oplus1},q\oplus1\rangle
+(|0,1,1,\overline{q},q\rangle)\}_{a,b,c,1,2,e}.
\end{equation}
When Alice sends his information qubits in next round, Eve can not
know these information qubits of Alice are in single particle
state or two qubit-state ahead of time. So Eve should not decide
how to operate and eavesdrop information.

Even if one of the legitimate users (Bob) is dishonest and wants
to eavesdrop Alice's information by intercept-and-resend attack
strategy, he can't also eavesdrop the information without being
detected. For example, after Alice sent two information qubits,
Bob intercepts the two qubits and sends his qubit 2' to charlie.
Charlie would take the qubit 2' for Alice's qubit 2. If Alice
sends two single particles as information qubits and entangles
Bob's qubit 1 to the carrier, she announces her operations after
confirming Bob and Charlie have received their qubits. Then Bob
can perform a CNOT gate $C_{b1}$ and two single particle
measurements on the qubits 1 and 2. So he can get the value of
$q_1\oplus q_2$. Bob also knows the value of $q_{2'}$, and then he
can announce the value of $q_{1'}$ according to $q_1\oplus
q_2=q_{1'}\oplus q_{2'}$. As a result Bob can get the secret
information without being detected. But when Alice entangles
Charlie's qubit 2 to the carrier, Charlie should perform a CNOT
gate $C_{c2'}$ and make a measurement on her receiving qubit 2'.
She will get the value of $q_{2'}$ or $q_{2'}\oplus1$. Bob can
still gain the value of $q_1\oplus q_2$ after knowing Alice's
operations. But he can't know the measurement result of Charlie is
$q_{2'}$ or $q_{2'}\oplus 1$. This would lead to 50\% difference
between the bit recovered by Charlie and the information bit Alice
has sent. If Bob intercepts the two qubits which are in two-qubit
state, he can't get the right information according to
Eq~(\ref{e6}). His eavesdropping behavior will be detected by the
legitimate users.

In summary, firstly, in our revised QSS protocol the secure
carrier can be used repeatedly to send secret information when
there is no presence of Eve. Secondly, Eve can't eavesdrop the
secret information by entangling her ancilla to the carrier. The
dishonest one of the two legitimate users can't gain information
without being detected. The revised protocol is unconditionally
secure. Thirdly, in the original protocol \cite{1} Alice encodes
information bits as a state $|q,q\rangle$ for sending odd bit. Bob
and Charlie can recover the information bit without each other's
assistance. In our protocol, Bob and Charlie must cooperate to
gain Alice's information in every round. So our protocol improves
the efficiency of the secret sharing. Furthermore, when Alice
sends $N$ bits information to Bob and Charlie, they will perform
$N$ Hadamard gates in the original protocol. In our protocol they
will perform Hadamard gate operation randomly. Our protocol
decreases the number of Hadamard gate operations.

\end{document}